\documentclass[12pt]{article}

\usepackage{amssymb}
\usepackage{amsmath}
\usepackage{amscd}
\usepackage{latexsym}
\usepackage{graphicx}

\usepackage{cite}
\newcommand{\be}{\begin{equation}}
\newcommand{\ee}{\end{equation}}

\newcommand{\dlt}{\delta}
\newcommand{\prt}{\partial}
\newcommand{\br}{{\bf r}}

\newcommand{\vp}{\varphi}
\newcommand{\ep}{\varepsilon}
\newcommand{\al}{\alpha}
\newcommand{\ra}{\rightarrow}
\newcommand{\sgm}{\sigma}

\newcommand{\Gm}{\Gamma}
\newcommand{\dgr}{\dagger}

\newcommand{\Lbd}{\Lambda}
\newcommand{\rgl}{\rangle}
\newcommand{\lgl}{\langle}

\begin{document}

\begin{center}
 
{\Large{\bf Quantum systems of atoms with \\ [2mm]

highly singular interaction potentials} \\ [5mm]

V.I. Yukalov$^{1,2}$ and E.P. Yukalova$^{3}$}  \\ [3mm]

{\it
$^1$Bogolubov Laboratory of Theoretical Physics, \\
Joint Institute for Nuclear Research, Dubna 141980, Russia \\ [2mm]

$^2$Instituto de Fisica de S\~ao Carlos, Universidade de S\~ao Paulo, \\
CP 369, S\~ao Carlos 13560-970, S\~ao Paulo, Brazil \\ [2mm]

$^3$Laboratory of Information Technologies, \\
Joint Institute for Nuclear Research, Dubna 141980, Russia } \\ [3mm]

Correspondence: Vyacheslav I. Yukalov (yukalov@theor.jinr.ru) \\

\end{center}

\vskip 1cm

\begin{abstract}
Quantum statistical systems, composed of atoms or molecules interacting with each other 
through highly singular non-integrable potentials, are considered. The treatment of such 
systems cannot start with the standard approximations such as Hartree, Hartree-Fock or
Hartree-Fock-Bogolubov approximations because of non-integrability of the interaction 
potentials leading to divergences. It is shown that the iterative procedure for Green 
functions can be rearranged so that the starting approximation takes into account 
regularizing atomic correlations. Then all the following approximation orders contain 
only the regularized interaction potential producing no divergences. The method of 
constructing the regularizing correlation function is suggested, based on the solution 
of the scattering equation in the form of asymptotic series at short distance, which 
can be extrapolated to arbitrary spatial variables by means of self-similar approximation 
theory. Regularizing correlation functions for several kinds of atomic systems are 
exemplified.  
   
\end{abstract}

\vskip 2cm
{\parindent=0pt
{\it Keywords}: highly singular potentials, correlated iterative procedure, 
regularizing correlation functions, regularized interaction potentials 
}

\newpage

\section{Introduction}

For treating quantum statistical systems with integrable interaction potentials, there exists 
the standard procedure of solving the system of equations for reduced density matrices 
\cite{Davidson_1,Coleman_2} or Green functions \cite{Kadanoff_3,Bonch_4,Yukalov_5}, starting 
an iterative process either with the approximation for free atoms or at the mean-field level 
and continuing the process until the required or achievable order. However for atomic systems 
with non-integrable interaction potentials this procedure does not work because of arising
divergences. Then one often limits oneself by an effective mean-field approximation where 
the bare non-integrable potential is replaced by a regularized potential that is integrable.     

For example, in the theory of cold quantum gases with low density, where the interaction 
radius is much smaller than the mean distance between atoms, one employs the effective 
interaction potential represented by the delta-function proportional to scattering length, 
which makes the potential integrable. Then as a working approximation one can choose, 
depending on the considered situation, a mean-field approximation of the Bogolubov or 
Hartree-Fock-Bogolubov form \cite{Dalfovo_1999,Courteille_2001,Andersen_2004,Yukalov_2004,
Bongs_2004,Yukalov_2005,Posazhennikova_2006,Yukalov_2007,Proukakis_2008,Yurovsky_2008,
Yukalov_2011,Yukalov_2016,Yukalov_2023,Yukalov_2024}. 

However, an important question remains unclear. How is it legitimate to use the same 
effective interaction potential in the following higher-order iterative steps, especially
when one needs to describe a dense system of atoms? Is it admissible to employ the 
effective-potential representation in higher-order approximations or the bare 
interaction potential has to be used in these higher-order approximations? In the latter 
case, the problems caused by the non-integrability of the bare interaction potential would 
reappear again.   

A potential $V({\bf r})$ is called highly singular when it is not integrable, such that
$$
\left| \; \int V(\br) \; d\br \; \right| \; \ra \; \infty \;   .
$$
For example, the potentials of power law $V({\bf r}) \propto 1/r^n$, with $n > d$, where 
$d$ is spatial dimensionality, upon integration, give the short-range terms diverging at 
$r \ra 0$. Therefore, it is impossible to start an iterative procedure from the Hartree, 
Hartree-Fock, or Hartree-Fock-Bogolubov approximations, since the Hartree potential 
$$
\left| \; \int V(\br-\br') \; \rho(\br') \; d\br' \; \right| \; \ra \; \infty 
$$
diverges for highly singular interactions.  

At the same time, the atomic and molecular systems with highly singular interaction 
potentials are widespread. The standard method of dealing with such systems is to resort,
from the very beginning, to second-order correlation functions defined by the scattering
matrix or variational conditions \cite{Bogolubov_6,Dupuis_7}. Such an approach is rather 
complicated and usually, is accepted as a single-step approximation. However, it would be 
highly desirable to develop a procedure, similar to that used for systems with integrable 
potentials, where one could start from a not too complicated approximation, say of a 
mean-field type, and proceed further having a well prescribed way of defining any desired 
higher-order approximation.    

To realize the above program requires to solve two problems. First, it is necessary to 
prove that, when introducing at the initial step an effective regularized potential, then 
in all higher-order approximations the same regularized potential would be present, so that 
no divergences would arise, despite the bare interactions through highly singular potentials. 
Second, as is clear, the desired initial approximation should somehow include interparticle 
correlations. How would it be possible to define this initial approximation so that not
to make it too complicated at the same type preserving the physics of the system? In the 
present paper, we suggest a resolution of these problems.   

Throughout the paper, the system of units is accepted where the Planck constant $\hbar$
is set to one.

\section{Iterative Procedure}

The possibility of developing an iterative procedure for Green functions so that, starting 
with an initial approximation enjoying the desired properties, one could define all following 
approximation orders containing no divergences, has been shown in 
\cite{Yukalov_2004,Yukalov_8}. In the situation of interest, the initial approximation 
has to be based on a regularized potential enjoying a nonsingular smooth form. To make the 
present paper self-consistent, let us briefly remind the main steps of the iterative 
procedure, keeping in mind a quantum system without gauge symmetry breaking.    

We use the techniques of Green functions \cite{Kadanoff_3,Bonch_4,Yukalov_5} that are 
applicable at zero as well as at finite temperatures. Let $x$ be a set of the spatial 
variable ${\bf r}$ and internal degrees of freedom, like spin, if any. For compactness, 
we accept the abbreviated notation for the dependence of functions:
\be
\label{1}
 f(12\ldots n) \; \equiv \; f(x_1,t_1,x_2,t_2,\ldots,x_n,t_n) \;  .
\ee
For instance, the Dirac delta function is
\be
\label{2}
\dlt(12) \; \equiv \dlt(x_1 - x_2) \; \dlt(t_1 - t_2) \;   .
\ee
The differentials are denotes as
\be
\label{3}
 d(12\ldots n) \; \equiv \; dx_1 \; dt_1 \; dx_2 \; dt_2 \ldots dx_n \; dt_n \; .
\ee
In this picture, the interactions are treated as retarded, such that
\be
\label{4}
V(12) \; \equiv V(x_1,x_2) \; \dlt(t_1 - t_2 + 0) \;  .
\ee

We shall use the causal Green functions, the single-particle function
\be
\label{5}
G(12) \; = \; - i \; \lgl \; \hat T \; \psi(1) \; \psi^\dgr(2) \; \rgl \; ,
\ee
where ${\hat T}$ is a chronological operator, and $\psi$ is a field operator, and 
the two-particle Green function
\be
\label{6}
G(1234) \; = \; -  \;
\lgl \; \hat T \; \psi(1) \;\psi(2) \; \psi^\dgr(3)\; \psi^\dgr(4) \; \rgl \;   .
\ee

Green functions are intimately related to reduced density matrices \cite{Yukalov_5}, 
thus the first-order density matrix is
\be
\label{7}
\rho(x_1,x_2,t) \; = \; \pm i \lim_{t_i\ra t} G(12) \qquad ( t_2 > t_1) \;   ,
\ee
where the upper sign is for the Bose statistics and the lower, for the Fermi statistics,
while the second-order density matrix is
\be
\label{8}
 \rho_2(x_1,x_2,x_3,x_4,t) \; = \; - \lim_{t_i\ra t} G(1234) \qquad
 ( t_4 > t_3 > t_2 > t_1) \;   .
\ee
Technically, it seems to be more convenient to work with Green functions. After finding 
the Green functions it is always possible to pass, if necessary, to reduced density 
matrices (\ref{7}) and (\ref{8}). 

We keep in mind the Hamiltonian of the standard form
$$
H \; = \; \int \psi^\dgr(x) \; \left[ \; - \;
 \frac{\nabla^2}{2m} + U(x) - \mu(x) \; \right] \; \psi(x) \; dx \; +
$$
\be
\label{9}
+ \; \frac{1}{2}  
\int \psi^\dgr(x) \; \psi^\dgr(x') \; V(x,x') \; \psi(x') \; \psi(x) \; dx dx' \;,
\ee
in which $U(x)$ is an external potential and $\mu(x)$ is a local chemical potential. The 
interaction potential $V(x,x')$ is assumed to be strongly singular, being non-integrable.  

The equation of motion for the single-particle Green function (\ref{5}) can be written as
\be
\label{10}
\int G^{-1}(13) \; G(32) \; d(3) \; = \; \dlt(12) \;  ,
\ee
where the inverse Green function is 
\be
\label{11}
G^{-1}(12) \; = \; \left[ \; 
i \; \frac{\prt}{\prt t_1} + \frac{\nabla^2}{2m} - U(1) + \mu(1) \; \right] \; \dlt(12) - 
\Sigma(12)
\ee
and the self-energy is 
\be
\label{12}
\Sigma(12) \; = \; \pm i \int V(13) \; G_2(1334) \; G^{-1}(42) \; d(34) \;   .
\ee

The equation of motion for the two-particle Green function (\ref{6}) can be obtained by
involving the Schwinger representation \cite{Kadanoff_3,Martin_10}
\be
\label{13}
 G_2(1223) \;  = \; G(13) \; G(22) \mp \frac{\dlt G(13)}{\dlt\mu(2)} 
\ee
and varying equation (\ref{10}), which gives
$$
G_2(1223) \;  = \; G(13) \; G(22)  \pm G(12) \; G(23) +
$$
\be
\label{14}
 +
\int G(14) G(53) \; \frac{\dlt\Sigma(45)}{\dlt G(67)} \;
\left[ \; G_2(6227) - G(67) G(22) \; \right] \; d(4567) \;  .
\ee

In the standard approach, the equation for the two-particle Green function is solved by
means of an iterative procedure starting with the Hartree-Fock approximation corresponding 
to the first two terms in the right-hand side of Eq. (\ref{14}),
\be
\label{15}
  G_{HF}(1223) \;  = \; G(13) \; G(22)  \pm G(12) \; G(23) \;  .
\ee
But, as is mentioned above, in the case of strongly singular potentials, such a way is 
not admissible because of the divergences occurring at the Hartree-Fock level. However, 
the choice of an initial approximation for an iterative procedure, strictly speaking, 
can be varied and in the present situation it is necessary to choose it so that to avoid 
divergences. This can be done by starting an iterative procedure not from the pure 
Hartree-Fock form, but with a form taking into account particle correlations. Defining 
the regularizing correlation function $g(12)$, we would like to accept as an initial 
approximation the correlated Hartree-Fock form
\be
\label{16}
 G_2^0(1223) \; = \; g(12) \; G_{HF}(1223) \;  .
\ee

Introducing the doubling function $D(123)$  by the relation
\be
\label{17}
G_2(1223) \; = \; g(12) \int D(124) G(43) \; d(4) \;   ,
\ee
we get for the self-energy (\ref{12}) the expression
\be
\label{18}
 \Sigma(12) \; = \; \pm i \int \Phi(13) D(132) \; d(3) \;  ,
\ee
in which 
\be
\label{19}
\Phi(12) \; = \; g(12) V(12)
\ee
is a regularized interaction potential. The regularizing correlation function should make 
the regularized potential integrable, such that
\be
\label{20}
 \left| \; \int \Phi(12) \; d(2) \; \right| \; < \; \infty \; .
\ee
   
In order that the initial approximation for the two-particle Green function (\ref{17})
would have the form (\ref{16}), the initial approximation for the doubling function
should be
\be
\label{21}
 D_0(123) \; = \; \dlt(13)  G(22) \pm G(12) \dlt(23) \;  .
\ee
Let us introduce an operator ${\hat X}(\Sigma)$ whose action on a three-point function 
$f(123)$ reads as
$$
 \hat X(\Sigma) \; f(123) \; = \; [ \; 1 - g(12) \; ] \; f(123) \; +
$$
\be
\label{22}
+
\int G(14) \; \frac{\dlt\Sigma(43)}{\dlt G(56))} \;
\left[\; g(12) \int f(527) G(76) \; d(7) - G(56) G(22) \; \right] d(456) \; .
\ee
Then equations (\ref{14}) and (\ref{17}), by straightforward, although cumbersome, 
rearrangements can be reduced to the iterative relation
\be
\label{23}
D_{n+1}(123) \; = \; D_n(123) + \hat X^{n+1}(\Sigma_{n+1}) \; D_0(123) \;   ,
\ee
with $D_n$ defining $\Sigma_{n+1}$ by equation (\ref{18}).

In this way, starting from the correlated Hartree-Fock approximation, we have a well 
defined iterative procedure allowing us to obtain any subsequent approximations. Then
the self-energy in the first approximation order becomes
\be
\label{24}
\Sigma_1(12) \; = \; 
\dlt(12) \int \Phi(13) \; \rho(3) \; d(3) + i \Phi(12) G(12) \; .
\ee
In the second order, the main part of the self-energy reads as
\be
\label{25}
\Sigma_2(12) \; = \;  \Sigma_1(12) - \int \Phi(13) G(43) \Gm(1324)\; d(34) \;  ,
\ee
with the vertex
\be
\label{26}
\Gm(1234)  \; = \; G(14) G(23) \Phi(43) \pm G(13) G(24) \Phi(34) \;  .
\ee
The correcting terms, proportional to $g(1-g)$, are omitted here, assuming that these 
correcting terms are small, which can be achieved by the choice of the regularizing 
function $g(12)$. If necessary, these terms can be taken into account \cite{Yukalov_8}.

As is seen, we have managed to realize a procedure, where at each step there occurs
only the regularized potential $\Phi$ defined in (\ref{19}), but not the bare strongly 
singular potential $V(r)$. At all steps of the iterative procedure the interaction 
potential enters through the regularized expression (\ref{19}), thus containing no 
divergences, provided the regularizing correlation function $g(12)$ is chosen 
appropriately. This choice is the main content of the following section.

\section{Correlation function}

The problem of defining correlation functions is a rather old one. Considering the 
singular behavior of the interaction potentials in real space, we shall deal with the
functional dependence on the spatial variable ${\bf r}$ whose absolute value is
$r = |{\bf r}|$. For an equilibrium system, there is no dependence on time. The 
straightforward understanding could be to consider, as $g({\bf r})$, the pair correlation 
function
\be
\label{27}
 \frac{\rho_2(\br_1,\br_2,\br_1,\br_2)}{\rho(\br_1)\rho(\br_2)} \;  .
\ee
However this function is not known and to find it would constitute a separate rather 
complicated task. Moreover, we do not need an exactly defined pair correlation function. 
What we need is to model an approximate regularizing correlation function that would 
allow us to avoid divergences connected with the strongly singular interaction 
potentials, while treating the iterative scheme for Green functions.

Kirkwood \cite{Kirkwood_11} suggested to define the correlation function from the 
experiment by connecting it to the structure factor that can be measured. Some other 
semi-phenomenological forms of the correlation function have also been discussed 
\cite{Yukalov_12}. However, our aim is to define all quantities from theory, without 
resorting to phenomenological palliatives.    

It seems, the direct way for describing correlations between particles could be by 
addressing the scattering equation \cite{Bogolubov_6}
\be
\label{28} 
 \frac{d^2\chi(r)}{dr^2} \; - \; m \; V(r) \; \chi(r) \; = \; 0 \;  ,
\ee
with the asymptotic condition 
\be
\label{29}
\chi(r) \; \ra \; 1 \qquad ( r \ra \infty) \;   .
\ee

For convenience, let us pass to dimensionless quantities. Suppose, there exist typical
parameters characterizing the length $\sigma$ and energy $\varepsilon$ of atomic
interactions. Then we define the dimensionless interaction potential $v(x)$ as a function 
of the dimensionless variable $x$, such that
\be
\label{30}
v(x) \; \equiv \; \frac{V(r)}{\ep} \; , \qquad
x \; \equiv \; \frac{r}{\sgm} \;   .
\ee
The dimensionless solution of the scattering equation (\ref{28}) is denoted as
\be
\label{31}
\vp(x) \; \equiv \; \chi(\sgm x) \;   .
\ee
Then the equation becomes
\be
\label{32}
 \frac{d^2\vp}{d x^2}  \; - \; \frac{v}{\Lbd^2}\; \vp \; = \; 0 \;  ,
\ee
where 
\be
\label{33}
\Lbd \; \equiv \; \frac{1}{\sgm\; \sqrt{m\ep} } 
\ee
is a dimensionless parameter.   

Our aim is to find a sufficiently general analytic expression for the correlation 
function that could be used for different interaction potentials. Recall that we keep
in mind a strongly singular potential that at short distance diverges as 
$V(r) \propto r^{-n}$, with $n > d$. Keeping in mind the three-dimensional space, we
have $n > 3$. The potential at $r \ra 0$ is assumed to be repulsive, so that there are 
no bound states. In that way, our aim is to find an approximate solution of the scattering
equation (\ref{32}).

Let us introduce the quantity
\be
\label{34}
 s(x) \; \equiv \; {\rm sgn}\; v(x)  \sqrt{x^n \; |\; v(x) \; | } \;  .   
\ee
Then the dimensionless interaction potential reads as
\be
\label{35}
 v(x) \; = \; \frac{1}{x^n} \; s^2(x) \;   .
\ee
And equation (\ref{32}) becomes
\be
\label{36}
 \frac{d\vp^2}{d x^2} \; - \; \frac{s^2}{\Lbd^2 x^n} \; \vp \; = \; 0 \;  .
\ee

Since the factor $1/x^n$, at short distance, is assumed to be the fast varying part 
of the potential, then the expression $x^n v(x)$ is smoother than $1/x^n$. More 
exactly, at small $x$, the quantity $s^2$ varies slower than $1/x^n$, so that
\be
\label{37}
\left| \frac{d s^2}{dx} / \frac{d}{dx}\left( \frac{1}{x^n}\right) \; \right| \; \ll \; 1 
\qquad ( x\ra 0) \;   .
\ee
Therefore, defining an approximate solution of (\ref{36}) at small $x$, we can keep the 
quantity $s$ as a quasi-integral of motion with respect to $r$. This separation on fast 
and slow variables is very useful in solving differential equations by the method of scale 
separation, or averaging method \cite{Bogolubov_14,Nayfeh_15}. The idea is to find 
a solution at small $x$ and then extrapolate this solution to arbitrary values of the 
variable $x$ by employing self-similar approximation theory. 

The solution of equation (\ref{36}), under fixed $s$, reads as
\be
\label{38}
\vp(x) \; = \; \sqrt{x} \; Z_\mu(z) \qquad 
\left( \mu \equiv - \; \frac{1}{n-2} \right) \;   ,
\ee
where the notation for the variable $z$ is
\be
\label{39}
z \; = \; - \; \frac{i}{x^{n/2-1}} \; \frac{2s}{(n-2)\Lbd} \;   .
\ee
Here $Z_\mu$ is a cylindrical function
\be
\label{40}
Z_\mu(z) \; = \; C_1 J_\mu(z) + C_2 N_\mu(z) \;   ,
\ee
in which $J_\mu$ is the Bessel function of the first kind and $N_\mu$ is the Bessel 
function of the second kind (Neumann function).   

Employing the properties of the Bessel functions \cite{Jahnke_20}, at small $x$ we have
\be
\label{41}
\vp(x) \; \simeq \; C x^{n/4} \sum_{p=0}^k a_p x^{(n/2-1)p} \;
\exp\left\{ - \; \frac{2s\; x}{(n-2)\Lbd x^{n/2}} \right\} \;   ,
\ee
with the coefficients
$$
a_p \; = \; \frac{\Gm(\mu+p+1/2)}{p! \; \Gm(\mu-p+1/2)} \;
\left[ \; \frac{(n-2)\Lbd}{4s(0)} \; \right]^p \;  .
$$
   
A series in powers of a parameter $x \ra 0$, derived for the asymptotically small 
parameter, can be extrapolated to the arbitrary values of this parameter by means of 
self-similar approximation theory 
\cite{Yukalov_21,Yukalov_22,Yukalov_23,Yukalov_24}. We shall not expose here the overall 
theory, but in Appendix we present the main results allowing us to realize the extrapolation. 
The self-similar extrapolation becomes precisely defined when the limiting condition for 
large $x \ra \infty$ is available. In our case, we do have such a large-variable limit 
requiring that
\be
\label{42}
|\; \vp(x) \; | \; \ra \; 1 \qquad ( x \ra \infty) \;    .
\ee
Then the series in Eq. (\ref{41}), in front of the exponential, can be summed by 
resorting to the self-similar approximation theory with the use of self-similar factor 
approximants \cite{Yukalov_26,Gluzman_27,Yukalov_28}. Employing the first two terms 
of the series,
we find
\be
\label{43}
\vp(x) \; = \; \frac{A^{\al/2} x^{n/4}}{(1+A x^{n/2-1})^{\al/2}} \; 
\exp\left\{ - \; \frac{2sx}{(n-2)\Lbd x^{n/2} } \right\} \;   ,
\ee
where 
$$
A \; = \; \frac{(n-2)\Lbd \; \Gm(\mu+3/2)}{2s(0)\;\Gm(\mu+1/2)\;\al} \;
\left( \frac{1}{2} - \mu \right) \; , \qquad
\al \; = \; \frac{n}{n-2} \;  .
$$

The sought correlation function reads as
\be
\label{44}
 g(x) \; = \; | \; \vp(x) \; |^2 \qquad \left( x = \frac{r}{\sgm} \right) \;  ,
\ee
which gives
\be
\label{45}
 g(x) \; = \; \frac{A^\al x^{n/2}}{(1+A x^{n/2-1})^\al} \; 
\exp\left\{ - \; \frac{4x\; {\rm sgn}\; v(x)}{(n-2)\Lbd} \;
\sqrt{|\; v(x) \; | } \; \right\} \;  .
\ee

As an example of a typical strongly singular potential let us consider the Lennard-Jones
potential
\be
\label{46}
 V(r) \; = \; 4\ep \; \left[ \; \left( \frac{\sgm}{r}\right)^{12} - 
\left( \frac{\sgm}{r}\right)^6 \; \right] \;   .
\ee
Quite a large number of atoms and molecules interact through this potential, because of 
which it is widely used in statistical physics, quantum chemistry, and molecular physics 
\cite{Stone_29,Lenhard_30}. In dimensionless units, we have
\be
\label{47}
 v(x) \; = \; 4 \left( \frac{1}{x^{12}} \; - \; \frac{1}{x^6} \right) \;  .
\ee
For the parameters entering function (\ref{45}), we find $\mu = -0.1$, $A = 1/2$, and
$\alpha = 6/5$. Then the correlation function (\ref{45}) becomes
\be
\label{48}
g(x) \; = \; \frac{A^{6/5} x^6}{(1+Ax^5)^{6/5}} \;
\exp\left\{ - \; \frac{2x\;{\rm sgn}\; v(x)}{5\Lbd} \;
\sqrt{|\; v(x)\; | } \; \right\} \;   .
\ee

In the dimensionless units we employ, atomic systems are characterized by the parameter
$\Lambda$. The correlation function (\ref{48}) is shown in Fig. 1 for Helium isotopes,
$^3$He ($\Lambda = 0.494$), $^4$He ($\Lambda = 0.430$), and $^6$He ($\Lambda = 0.347$), 
Figure 2 shows the correlation function for polarized Hydrogen ($\Lambda = 0.740$),
polarized Deuterium ($\Lambda = 0.523$), and polarized Tritium ($\Lambda = 0.428$).
The regularized potential (\ref{19}) becomes smoothed by the correlation function
and nonsingular, as is seen from Fig. 3 and Fig. 4 for the corresponding materials 
characterized by the parameter $\Lambda$.    

%Figure 1
\begin{figure}[ht]
\centerline{
\includegraphics[width=9cm]{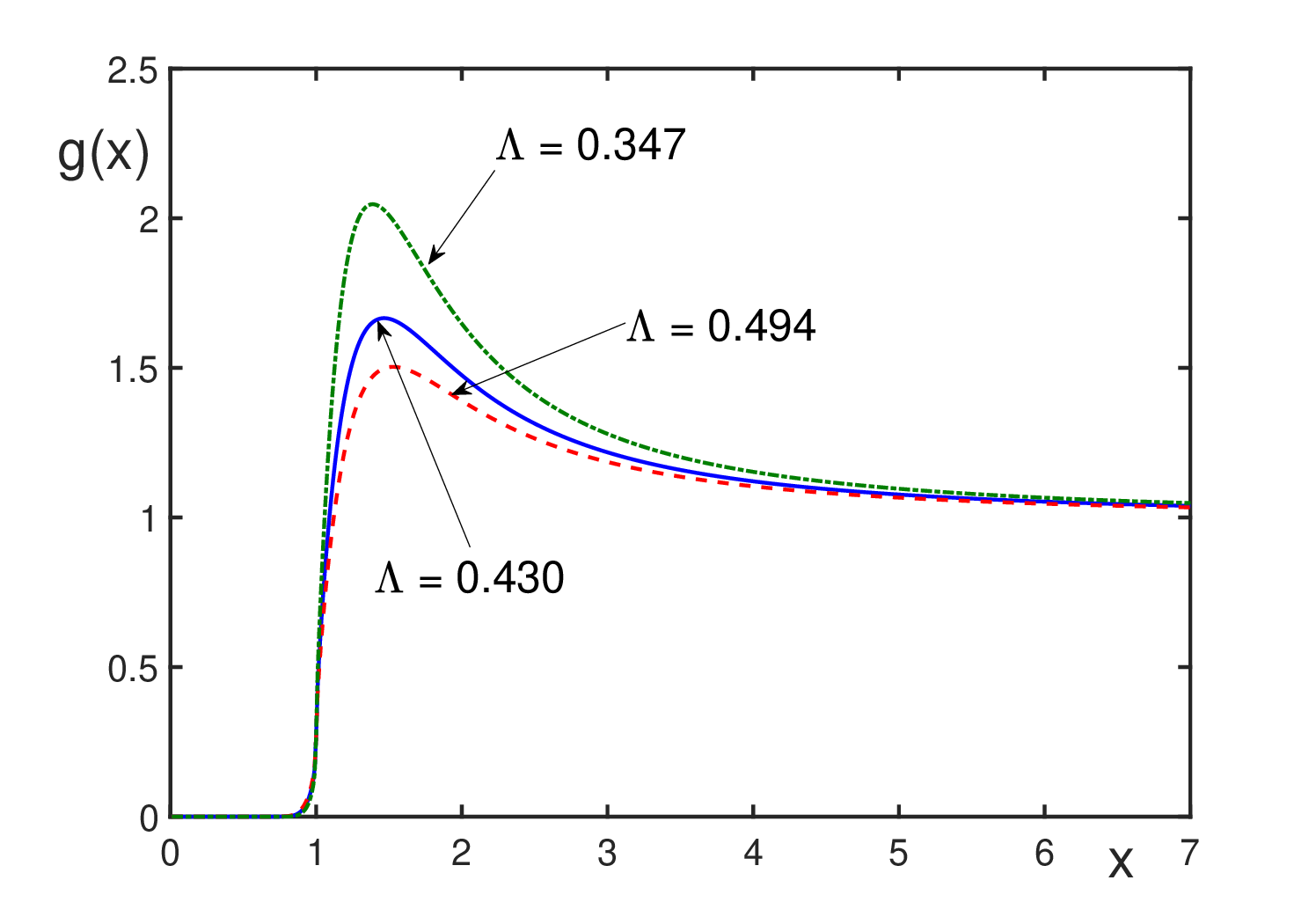}  }
\caption{\small
Regularizing correlation functions for Helium isotopes,  $^3$He ($\Lbd= 0.494$), 
$^4$He ($\Lbd=0.430$), and $^6$He ($\Lbd=0.347$).
}
\label{fig:Fig.1}
\end{figure}

%Figure 2
\begin{figure}[ht]
\centerline{
\includegraphics[width=9cm]{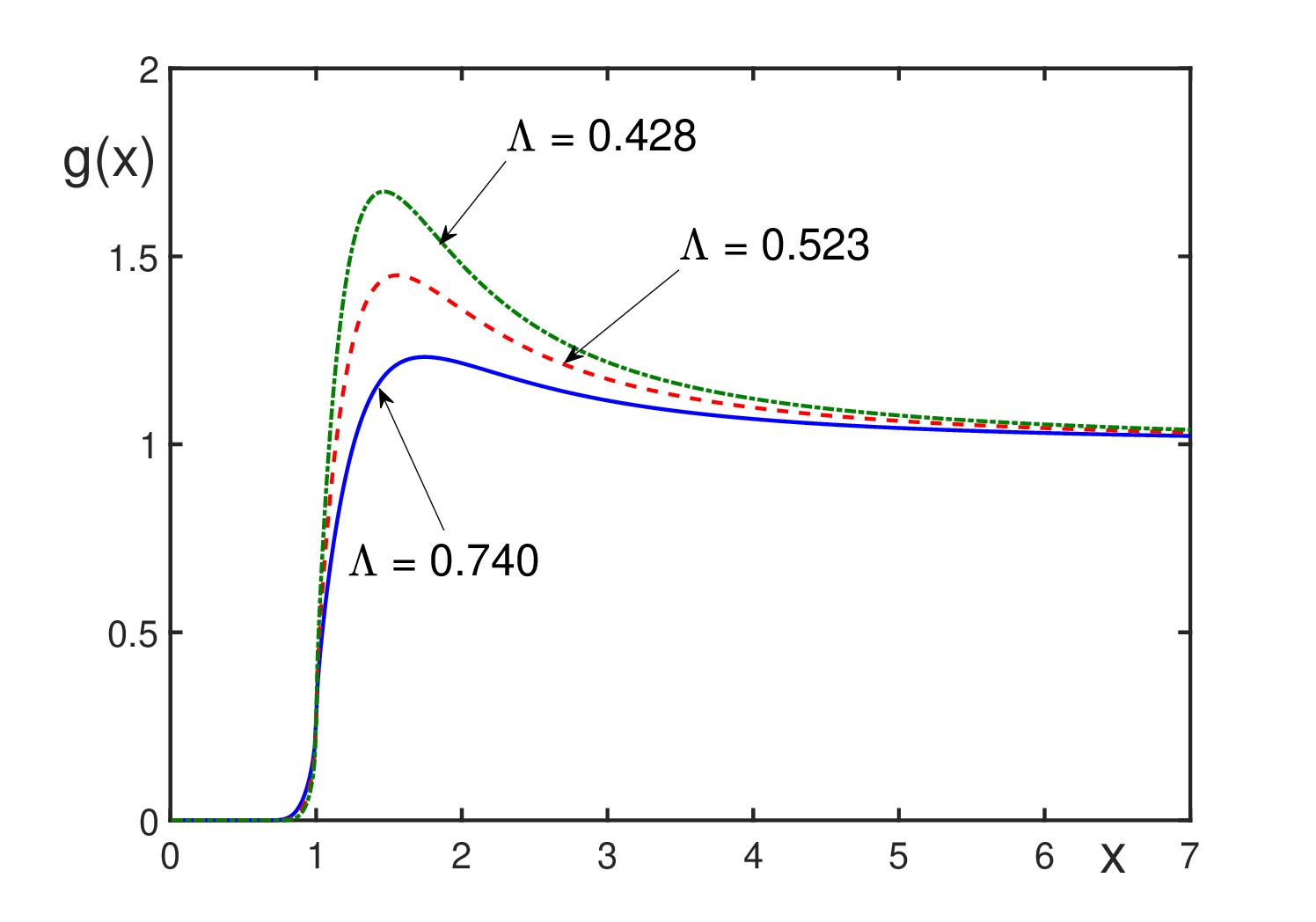}  }
\caption{\small
Regularizing correlation functions for polarized Hydrogen ($\Lbd=0.740$),
polarized Deuterium ($\Lbd=0.523$), and polarized Tritium ($\Lbd=0.428$).
}
\label{fig:Fig.2}
\end{figure}

%Figure 3
\begin{figure}[ht]
\centerline{
\includegraphics[width=9cm]{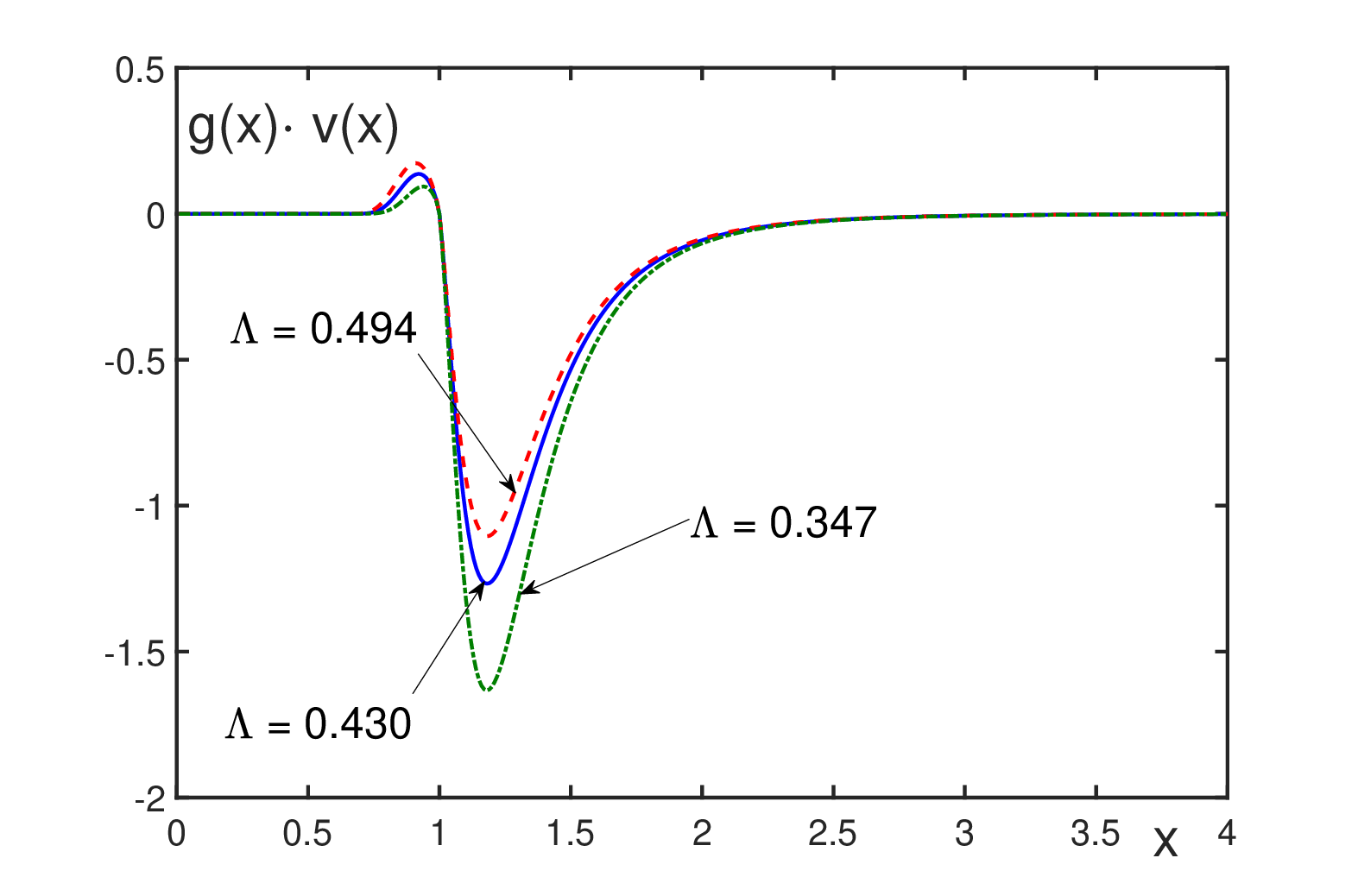}  }
\caption{\small
Regularized interaction potential for Helium isotopes,
$^3$He ($\Lbd=0.494$), $^4$He ($\Lbd=0.430$), and $^6$He ($\Lbd=0.347$).
}
\label{fig:Fig.3}
\end{figure}

%Figure 4
\begin{figure}[ht]
\centerline{
\includegraphics[width=9cm]{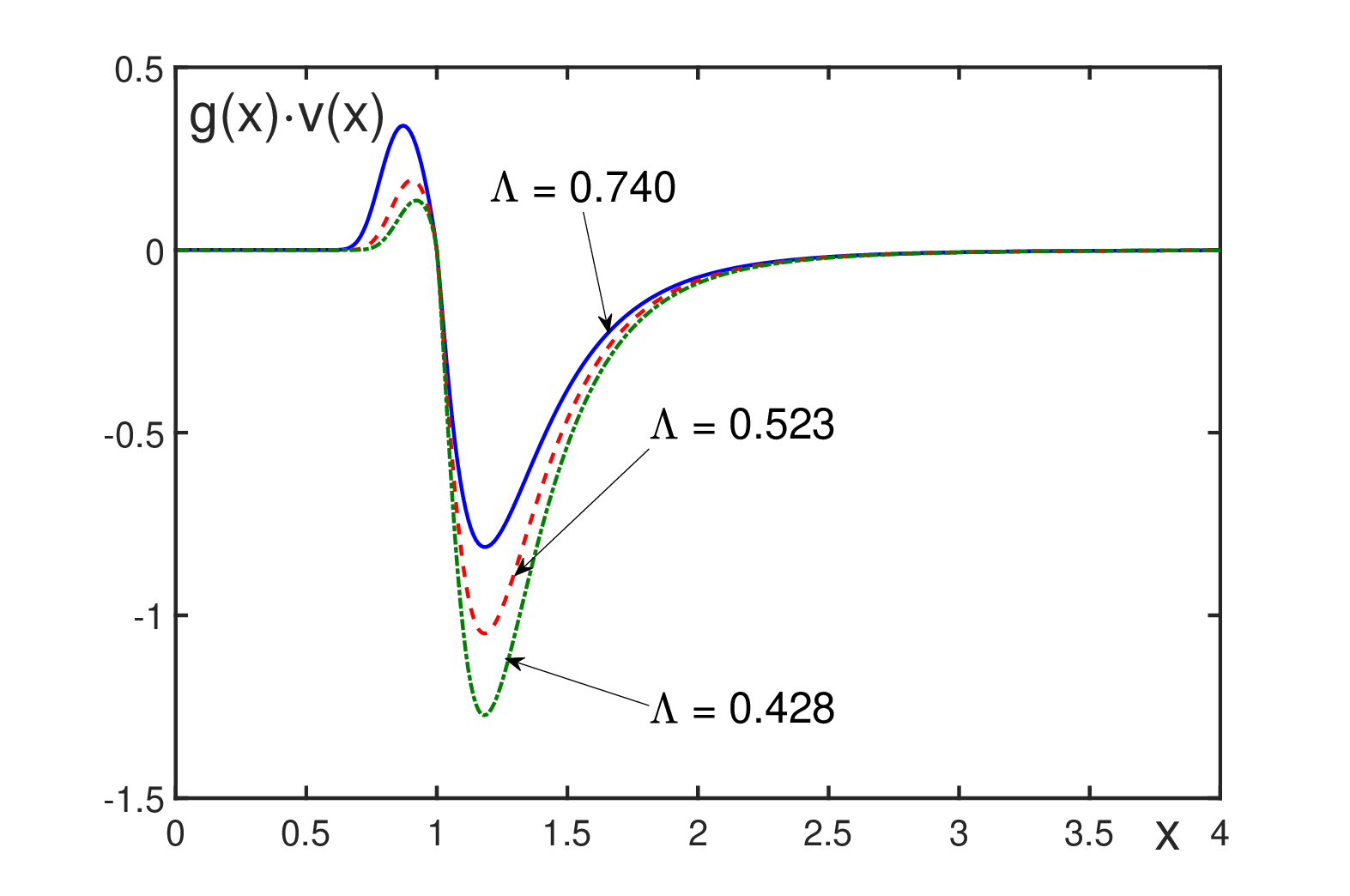}  }
\caption{\small
Regularized interaction potential for polarized Hydrogen ($\Lbd=0.740$),
polarized Deuterium ($\Lbd=0.523$), and polarized Tritium ($\Lbd=0.428$).
}
\label{fig:Fig.4}
\end{figure}

\section{Conclusion}

The goal of this paper has been twofold. First, to show that in principle an iterative 
procedure for statistical systems can always be rearranged so that one could start it
with an initial approximation enjoying the desired properties. The main attention is 
paid to the case of atoms interacting through strongly singular potentials. Because
of non-integrability of such potentials, the iterative procedure cannot start from a
Hartree-Fock type approximation. In order to smooth a highly singular interaction 
potential, one has from the very beginning to take into account the existence of 
particle correlations, whose presence regularizes the singular interaction potential. 

Defining a regularizing correlation function is another important problem. From one side, 
the regularizing correlation function does not need to, and cannot, be defined exactly 
as a pair correlation function that is not known. From the other side, this regularizing 
function has to possess the properties close to the pair correlation function in order 
to efficiently regularize the singular interaction potential making the regularized 
potential smooth. It is also very important that the regularizing function be of simple 
form, although being realistic, and valid for arbitrary interaction potentials. A method 
of constructing such regularizing correlation functions is suggested. The method is based 
on solving the scattering equation at short distance and then extrapolating this solution 
to arbitrary variables by means of the self-similar approximation theory satisfying the 
required boundary conditions. The examples of regularizing correlation functions for 
several quantum atomic substances are illustrated. The related regularized interaction 
potentials are shown to be smooth and nonsingular. 

It is important to stress that the developed iterative procedure in the first-order gives 
exactly the mean-field approximation, but with a regularized smooth interaction potential.
This regularized potential enters all higher approximation orders, so that no divergences 
occur. Starting form the second order, there appear corrections containing the powers of
$g(1-g)$, however the main terms retain their standard structure, but with the regularized
interaction potential. 

The developed approach can be useful for treating atomic and molecular systems with highly 
singular non-integrable interaction potentials. The approach gives a justification for the 
use of regularized interaction potentials, showing that, starting from the mean-field level,
the same regularized potential enters all higher-order approximations. The general form of 
the regularizing correlation function (\ref{45}) allows for its direct use for different
highly singular potentials diverging at short distance as $r^{-n}$, with $n > 2$.             
    
\vskip 2mm

\appendix
\section{Appendix. Self-similar approximation theory}

\numberwithin{equation}{section}
\setcounter{equation}{0}

The method we use in Sec. 3 for the extrapolation of an asymptotic series in powers of a 
small parameter to the arbitrary values of the parameter is based on the self-similar
approximation theory \cite{Yukalov_21,Yukalov_22,Yukalov_23,Yukalov_24}. We will not 
expand the overall theory here, but will just survey its main steps in order that the 
reader could grasp the idea why this method can provide accurate extrapolations from
small variables (or parameters) to arbitrary values of the variable, including its large
limiting values. A detailed description of all related mathematics can be found in the 
review article \cite{Yukalov_31}.  

Suppose we have derived an asymptotic series of a real-valued function 
\be 
\label{A.1}
f_k(x) \; = \; f_0(x) \sum_{n=0}^k a_n x^n \qquad (x \ra 0)
\ee
of a real variable $x \in [0, \infty)$, with $f_0(x)$ being a given function. We need to
extrapolate the series to arbitrary $x$ in the whole interval $[0, \infty)$. The basic 
idea explaining why this, in principle, is admissible is as follows. Although the truncated 
series $f_k(x)$ is derived for small $x \ra 0$, but its coefficients $a_n$ contain 
information on the whole sought function. This information can be extracted from the 
series by noticing how the approximants $f_k$ vary with the change of $k$. That is, it 
is necessary to find out the variation of $f_k$ with the change of $k$. And, knowing the 
variation of $f_k$ with $k$, it is possible to search for an effective extrapolation 
to large $k$. Technically, the following steps are required. 

\vskip 2mm

{\it Step 1}. {\it Implantation of control parameters}. The initial series consisting 
of the approximants $f_k$ are usually divergent, hence it is necessary to incorporate 
into it control parameters that could induce the series convergence. Denoting the set 
of control parameters by $u$, we transform the initial terms $f_k(x)$ into $F_k(x,u)$. 
The implantation of control parameters can be done, for instance, through the fractal 
transform
\be
\label{A.2}
F_k(x,\{ n_j\} ) \; = \; \prod_{j=1}^k x^{-n_j} ( 1 + b_j x) \;   ,
\ee
in which the control parameters are $n_j$ and $b_j$.        

\vskip 2mm
{\it Step 2}. {\it Specification of control functions}. The implanted control parameters
$u$ need to be converted into control functions $u_k(x)$ such that to induce the 
convergence of the sequence $\{F_k(x,u_k(x))\}$. Control functions can be defined either
by the minimization of a cost functional, or by training conditions. 

\vskip 2mm
{\it Step 3}. {\it Construction of approximation cascade}. The sequence $\{F_k(x,u_k(x))\}$
can be reformulated into a bijective sequence $\{y_k(f)\}$ representing the trajectory of
an approximation cascade, with  the approximation order playing the role of discrete time.
The cascade trajectory starts with the initial point $f = y_0(f)$. The motion of the 
cascade in the approximation space, in the vicinity of a fixed point, displays the property 
of self-similarity
\be
\label{A.3}
y_{k+p}(f) \; = \; y_k(y_p(f) ) \;   .
\ee
 
\vskip 2mm
{\it Step 4}. {\it Embedding cascade into flow}. The approximation cascade, that is a 
dynamical system in discrete time, can be embedded into an approximation flow that is
a dynamical system in continuous time, which is equivalent to considering the 
approximation order $k$ as a continuous variable $t$. The flow trajectory passes through
all points of the cascade trajectory, the property of self-similarity being preserved,
\be
\label{A.4}
y(t+t',f) \; = \; y(t,y(t',f) ) \;   .
\ee
  
\vskip 2mm
{\it Step 5}. {\it Searching for fixed points}. The self-similar relation (\ref{A.4}) can 
be transformed into the Lie differential equation
\be
\label{A.5}
\frac{\prt}{\prt t} \; y(t,f) \; = \; v(y)   
\ee
describing the evolution of the sequence terms, with $v(y)$ being the flow velocity. 
The fixed point $y_k^*$ of the evolution equation represents the effective limit of the 
approximation sequence $f_k^*(x)$, called the self-similar approximant of order $k$.

Accomplishing the above manipulations leads \cite{Yukalov_26,Gluzman_27,Yukalov_28}
to the self-similar factor approximants
\be
\label{A.6}
f_k^*(x) \; = \; f_0(x) \prod_{j=1}^{N_k} ( 1 + A_j x)^{n_j} \;   ,
\ee
where the control parameters $A_j$ and $n_j$ are prescribed by the training conditions
\be
\label{A.7}
 \lim_{x\ra 0} \; \frac{1}{n!}\; \frac{d^n}{dx^n} \left[\; 
\frac{f_k^*(x)}{f_0(x)}\; \right] \; = \; a_n 
\ee
and the boundary conditions. The number $N_k$, defining the number of factors in formula
(\ref{A.6}), equals the number of training conditions plus the number of boundary conditions. 

For example, let us know the large-variable boundary condition of the sought function
\be
\label{A.8}
 f(x) \; \simeq \; C x^\nu \qquad ( x \ra \infty) \;  ,
\ee
so that
\be
\label{A.9}
\lim_{x\ra 0} f_0(x) \; = \; f_0 \;   .
\ee
At the same time, expression (\ref{A.6}) at large $x$ yields
\be
\label{A.10}
 f_k^*(x) \; \simeq \; f_0 \prod_{j=1}^{N_k} A_j^{n_j} \; x^{\nu_k} \qquad 
( x \ra \infty) \;  ,
\ee
in which 
\be
\label{A.11}
\nu_k \; = \; \sum_{j=1}^{N_k} n_j \; .
\ee
Therefore the boundary conditions (\ref{A.8}) and (\ref{A.9}) give
\be
\label{A.12}
 f_0 \prod_{j=1}^{N_k} A_j^{n_j} \; = \; C \; , \qquad
\nu_k \; = \; \sum_{j=1}^{N_k} n_j \; = \; \nu \;  .
\ee
In the case considered in Sec. 3, we have the boundary condition (\ref{42}), hence  
$C = 1$ and $\nu = 0$, which results in expression (\ref{43}).

\section*{Funding} 

This research received no external funding.

\section*{Conflict of Interest Statement}

The authors declare no scientific or financial conflicts of interest.

\end{document}